\RequirePackage{fix-cm}
\documentclass[twocolumn,epjc3]{svjour3}  
\smartqed  
\RequirePackage{graphicx}
\journalname{Eur. Phys. J. C}
\usepackage{amsmath}
\usepackage{amsfonts}
\usepackage{amssymb}
\usepackage{multirow}
\usepackage{soul}
 
\def\NIM{Nucl.\,Instrum.\,Methods}

\def\Re{$^{187}$Re}
\def\Ho{$^{163}$Ho}
\def\Dy{$^{163}$Dy}

\def\Tr{$^3$H}

\def\de{$\Delta E$}
\def\fwhm{$_{\mathrm{FWHM}}$}
\def\Aec{$A_\mathrm{EC}$}

\def\fpp{$f_{pp}$}

\begin{document}
\title{Statistical sensitivity of \Ho\ electron capture neutrino mass experiments}
\author{Angelo Nucciotti\thanksref{e1,addr1,addr2}}
\thankstext{e1}{e-mail: angelo.nucciotti@mib.infn.it}
\institute{Dipartimento di Fisica ``G. Occhialini'', Universit\`a di Milano-Bicocca\label{addr1} \and	Istituto Nazionale di Fisica Nucleare, Sezione di Milano-Bicocca\label{addr2}}
\date{Received: date / Revised version: date}
%
%
\maketitle
\abstract{
Large calorimetric neutrino mass experiments using thermal detectors might be going to play a crucial role
in the challenge for assessing the neutrino mass. This paper describes a tool based on Monte Carlo methods which has been
developed to estimate the statistical sensitivity of calorimetric direct neutrino mass measurements using the \Ho\ electron capture decay. 
The tool is applied to investigate the effect of various experimental parameters. In this paper I report the results useful for
designing an experiment with sub-eV sensitivity.
%
} 
\section{Introduction}
\label{intro}
One of the challenges for particle physics in next decade will be to probe the neutrino absolute mass down to at least the lowest bound of the inverted hierarchy region, i.e. about 0.05\,eV \cite{Fogli}.

Present best limits on the neutrino absolute mass have been set using MAC-E filter spectrometers to analyze the end-point of \Tr\ beta decay \cite{MainzTroitsk} and are about 2\,eV. In a couple of years the new large MAC-E filter spectrometer of the KATRIN experiment will become operational with the aim to push the sensitivity to neutrino absolute mass down to about 0.2\,eV \cite{KATRIN}. 
With KATRIN, this experimental approach reaches its technical limits. It is therefore mandatory for the neutrino physics community to define alternative and complementary experimental methodologies to extend the reach of direct neutrino mass measurements.

The calorimetric measurement of nuclear decays with low end-point is a promising alternative way which has been already applied to \Re\ beta decay leveraging the powerful technique of low temperature calorimetry\,\cite{manu,mibeta,neutrino2010}.
More recently, the use of \Ho\ has been widely revived. Presently there are at least two projects 
working to perfom
high sensitivity experiments with \Ho: ECHo \cite{echo} and HOLMES, a follow up of the MARE project, which was recently funded by the European Research Council \cite{neutrino2010,galeazzi}. 
As one of the promoters of the HOLMES experiment, I developed a Monte Carlo code for assessing the statistical
sensitivity of calorimetric neutrino mass measurements based on \Ho\ electron capture (EC) decay. In this paper I collected
the most relevant results to share them with the growing community of scientists engaged in such type of experiments. 

\section{Calorimetric measurement of \Ho\ electron capture decay}
\label{calo}
In 1982 De Rujula and Lusignoli \cite{rujula83} discussed the calorimetric measurement of the \Ho\ spectrum as a mean for directly measuring the electron neutrino mass $m_\nu$. \Ho\ decays by electron capture (EC) to \Dy\ with a half life of about 4570 years and with the lowest known $Q$-value, which allows captures only from the M shell or higher. The decay $Q$-value has been experimentally determined only using the ratios of the capture probability from different atomic shells. The various determinations span from 2200 to 2800\,eV -- with a recommended value of $2555\pm16$\,eV \cite{Rei10} --, where the error is largely due to systematic uncertainties such as the errors on the theoretical atomic physics factors involved.

In a calorimetric EC experiment all the de-excitation energy is recorded. The de-excitation energy $E_c$ is the energy released by all the atomic radiation emitted in the process of filling the vacancy left by the EC decay, mostly electrons with energies up to about 2\,keV \cite{rujula83} (the fluorescence yield is less than $10^{-3}$). The calorimetric spectrum appears as a series of lines at the ionization energies $E_i$ of the captured electrons. These lines have a natural width $\Gamma_i$ of a few eV and therefore the actual spectrum is a continuum with marked peaks with Breit-Wigner shapes (Figure\,\ref{fig:spectra}). 
The spectral end-point is shaped by the same neutrino phase space factor $(Q-E_e)[(Q-E_e)^2-m_\nu^2]^{1/2}$ that appears in a beta decay spectrum, with the total de-excitation energy $E_c$ replacing the electron kinetic energy $E_e$. 
For a non-zero $m_\nu$, the de-excitation (calorimetric) energy $E_c$ distribution is expected to be
\begin{eqnarray}
\label{eq:E_c-distr}
 {d \lambda_\mathrm{EC}\over dE_c} &=& {G_{\beta}^2 \over {4 \pi^2}}(Q-E_c) \sqrt{(Q-E_c)^2-m_{\nu}^2} \;
\times  \\
&& \sum_i n_i  C_i \beta_i^2 B_i {\Gamma_i \over 2\,\pi}{1 \over (E_c-E_i)^2+\Gamma_i^2/4} \,, 
\nonumber
\end{eqnarray}
where $G_{\beta} = G_F \cos \theta_C$ (with the Fermi constant $G_F$ and the Cabibbo angle $\theta_C$), $E_i$ is the binding energy of the $i$-th atomic shell, $\Gamma_i$ is the natural width, $n_i$  is the fraction of occupancy, $C_i $ is the nuclear shape factor, 
$\beta_i$ is the Coulomb amplitude of the electron radial wave function (essentially, the modulus of the wave function at the origin) and $B_i$ 
is an atomic  correction for electron exchange and overlap. 

As for beta decay experiments, the neutrino mass sensitivity depends on the fraction of events close to the end-point, which in turn depends on the decay $Q$-value. In particular, the closer the $Q$-value to the highest $E_i$, the larger the resonance enhancement of the rate near the end-point, where the neutrino mass effects are relevant.

Because of the high specific activity of \Ho\ (about $2\times10^{11}$\,\Ho\ nuclei give one decay per second) the calorimetric measurements will be achieved by introducing relatively small amounts of \Ho\ nuclei in detectors whose design and physical characteristics -- i.e. material and size -- are driven almost exclusively by the detector performance requirements and by the de-excitation radiation containement \cite{echo,galeazzi}. 

\section{Monte Carlo simulation}
\label{sec:approach}
In this section we describe a frequentist Monte Carlo code developed to estimate the statistical neutrino mass sensitivity for a calorimetric
\Ho\ EC measurement\footnote{This is a frequentist Monte Carlo in the sense that, without making any {\it a-priori} hypothesis on the probability distribution of the measurement results (the neutrino mass squared), a large number of
toy experiments is performed and the frequency distribution of the results is considered. Since there is no true signal in the toy experiments, the sample mean is about 0 as expected and the sample standard deviation gives the instrumental statistical uncertainty which is defined as the instrumental sensitivity.
This approach has been checked against the sensitivity definition proposed in \cite{FeldmanCousins}, i.e. \textit{the average upper limit one would get from an ensemble of experiments with the expected background and no true signal}. Indeed the two approaches give similar -- though not identical -- results. However the definition in  \cite{FeldmanCousins} runs into problems when dealing with non-physical results (i.e. negative square neutrino masses). In fact fits of individual toy experiments may return a negative square neutrino mass and estimating the upper limit then requires an approach either Bayesian  or frequentist as described in \cite{FeldmanCousins}. On the contrary the approach used for the results reported in this paper does not require any further statistical ``trick'' to deal with the unavoidable non-physical results and it is therefore intrinsically robust.
}. The approach is a replica of the one outlined in \cite{sensbeta} for beta decay calorimetric experiments. 
It consists in the simulation of the spectra that would be measured by a large number of experiments carried out in a given configuration: the spectra are then fit as the real ones and the statistical sensitivity is deduced from the distribution of the obtained $m^2_\nu$ parameters.

This method proved to be extremely powerful since it allows to include all relevant experimental effects -- such as energy resolution, pile-up and background -- and also to estimate systematic uncertainties \cite{sensbeta}. In this paper however we will limit the discussion to the statistical sensitivity, since systematic effects in this kind of measurement are not fully known yet.

The parameters describing the experimental configuration are the total number of \Ho\ decays $N_{ev}$,
the FWHM of the Gaussian energy resolution \de \fwhm, the fraction of unresolved pile-up events $f_{pp}$ and the radioactive background $B(E)$.

The total number of events is given by \mbox{$N_{ev} = N_{det}A_\mathrm{EC}t_M$}, where $ N_{det}$, \Aec\ and  $t_M$ are the total number of detectors, the EC decay rate in each detector and the measuring time, respectively.

\begin{figure}[!tb]
\begin{center}
\resizebox{0.45\textwidth}{!}{\includegraphics*{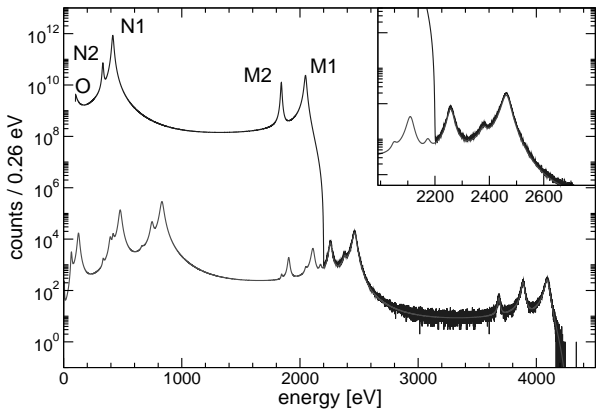} }
\end{center}
\caption{\label{fig:spectra} Full \Ho\ decay experimental spectrum simulated for $Q=2200$\,eV, $N_{ev}=10^{14}$, $f_{pp}=10^{-6}$,  $\Delta E_{\mathrm {FWHM}} = 2$\,eV, $m_\nu=0$. The bottom curve is a fit of the pile-up spectrum. The insert shows the end-point region of the spectrum.}
\end{figure}

Pile-up happens when two decays in one detector are too close in time and are mistaken as a single one with an apparent energy equal to the sum of the
two decays. In first approximation, this has a probability of $f_{pp}= \tau_R A_\mathrm{EC}$, where $\tau_R$ is the detector time resolution. The energy spectrum of pile-up events is given by the self-convolution of the calorimetric EC decay spectrum and extends up to $2Q$, producing therefore a background impairing the ability to identify the neutrino mass effect at the decay spectrum end-point $Q$.
In the case of \Ho\ decay, the pile-up events spectrum is quite complex and presents a number of peaks right at the end-point of the decay spectrum (Figure\,\ref{fig:spectra}).

The $B(E)$ function is usually taken as a constant $B(E)=bT$, where $b$ is the average background count rate for unit energy
and for a single detector, and $T=N_{det}\times t_M$ is the experimental exposure.

The theoretical spectrum $S(E_c)$ which is measured by the simulated toy experiments is then given by (Figure\,\ref{fig:spectra}):
\begin{eqnarray}
\label{eq:theory}
S(E_c) &=& [ N_{ev}(N_\mathrm{EC}(E_c,m_\nu)+f_{pp} \times \\ 
&& N_\mathrm{EC}(E_c,0)\otimes N_\mathrm{EC}(E_c,0))  +B(E_c) ]\otimes R_{\Delta E}(E_c)
\nonumber
\end{eqnarray}
where $N_\mathrm{EC}(E_c,m_\nu)$ is the \Ho\ spectrum as described by (\ref{eq:E_c-distr}) and with unity normalization, $B(E)$ is the background
energy spectrum, and the detector energy response function is given by 
\begin{equation}
\label{eq:gauss}
 R_{\Delta E}(E_c)=\frac{1}{\sigma \sqrt{2\pi}}e^{-\frac{E_c^2}{2\sigma^2}}
\end{equation}
with standard deviation $\sigma = \Delta E_{\mathrm {FWHM}}/2.35$.\footnote{In actual experiments $R_{\Delta E}(E_c)$ may have an explicit dependence on the energy $E_c$: usually the energy resolution \de\fwhm\ gets worse for increasing energy. This behaviour has not been included in the present investigation because it does not affect the experimental sensitivity, although it has to be considered when analysing real data to avoid systematic errors.}
For the simulations, the parameters $E_i$, $\Gamma_i$, $n_i$, $C_i$, $B_i$, and $\beta_i$ in (\ref{eq:E_c-distr}) are taken from \cite{Lus11}. 

The set of experimental spectra  -- typically between 100 and 1000 -- is obtained by 
fluctuating the spectrum $S(E_c)$  (\ref{eq:E_c-distr}) according to Poisson statistics.  Each simulated spectrum is then fitted using 
 (\ref{eq:E_c-distr}) and leaving $m^2_\nu$, $Q$, $N_{ev}$, $f_{pp}$ and $b$ as
free parameters. 
In a real experiment the atomic parameters describing the Breit-Wigner peaks will be determined from high statistics measurements. Still, a correct neutrino mass analysis of the experimental spectrum may require to leave some of them as free parameters in the fit, in particular the ones relative to the M1 peak. At the expenses of a much higher computing time, a modified version of the code has been developed to investigate the effect of this approach. Some tests have been carried out leaving the three additional M1 peak parameters free - i.e. position $E_{M1}$,  width $\Gamma_{M1}$, and relative intensity -- in few sample experimental configurations: the results show a worsening of the sensitivity however always well below 10\%.
The simulated experimental spectra are generated on an energy interval which is smaller than the full 0 -- $2Q$ interval and the fits are performed on sub-intervals of this. For most of the simulations presented in this paper, the fitting interval is 1500 to 3500\,eV. However, tests show that the results worsen less than 10\% by 
extending
the lower end of the fitting interval up to 2100\,eV -- i.e. to the right side of the M1 peak. 

The 90\% C.L. $m_\nu$ statistical sensitivity $\Sigma_{90}(m_\nu)$ of the simulated experimental 
configuration can be obtained from the distribution of the $m^2_\nu$ found by fitting the spectra. For a Gaussian distribution as the one found in the present work the sensitivity is then given by $\Sigma_{90}(m_\nu) = \sqrt{1.64 \sigma_{m_\nu^2}}$, where $\sigma_{m_\nu^2}$ is the standard deviation of the distribution:
\begin{equation}
\sigma_{m_\nu^2}^2 = \frac{1}{N-1} \sum_i (m_{\nu_i}^2 - \overline{m_\nu^2})^2 = \frac{N}{N-1} (\overline{m^4_\nu} - \overline{m^2_\nu}^2)
\end{equation}
where $N$ is the number of generated spectra and $m_{\nu_i}^2$ are the values found in each fit for $m^2_\nu$ fit parameter.

The statistical error on the sensitivity $\Sigma_{90}(m_\nu)$ is given by (see \cite{sensbeta} for details)
\begin{equation}
\label{eq:errorsigma}
\epsilon_{\Sigma_{90}(m_\nu)} = \frac{1.64}{2} \frac{\epsilon_{\sqrt{\overline y}}}{\Sigma_{90}(m_\nu)}
\end{equation}
where $y_i=(m_{\nu_i}^2 - \overline{m_\nu^2})^2$ and $\overline{y} \approx \sigma_{m_\nu^2}^2$.
Using equation (\ref{eq:errorsigma}) one finds that the statistical error on the Monte Carlo results is around 3\% and 1\% for about 100 and 1000 simulated experiments, respectively.

\subsection{Results}
\label{sec:results}
Given the large uncertainties on the \Ho\ EC $Q$-value, all the simulations have been performed for few $Q$-values picked in the interval 2200 -- 2800\,eV.

\begin{figure}[!tb]
\begin{center}
\resizebox{0.45\textwidth}{!}{\includegraphics*{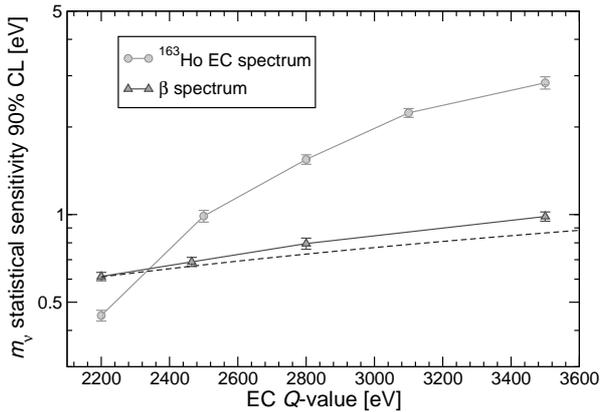}}
\end{center}
\caption{\label{fig:Q_dep} \Ho\ and beta decay experiments statistical sensitivity dependence on the $Q$-value for $N_{ev}=10^{12}$, \de\fwhm$=1$\,eV, $f_{pp}=0$, and $b=0$\,count/eV/s/detector.}
\end{figure}

First of all it is instructive to compare how the statistical sensitivity for a given statistics $N_{ev}$ depends on the $Q$-value 
in the case of \Ho\ and 
of a low energy beta decay with a spectral shape similar to the one of \Re\ \cite{sensbeta}. 
Figure\,\ref{fig:Q_dep} shows that \Ho\ experimental sensitivity depends on the $Q$-value more steeply than $Q^{3/4}$ (dashed line in Figure\,\ref{fig:Q_dep}) as for beta decays and, for $Q$-values smaller than about 2400\,eV, \Ho\ experiments are more favorable than beta decay ones. The details of the simulation are given in the caption of Figure\,\ref{fig:Q_dep}. The steeper behavior observed for \Ho\ decay is the result of the resonance enhancement caused by the proximity to the M1 capture peak.

\begin{figure}[!tb]
\begin{center}
\resizebox{0.43\textwidth}{!}{\includegraphics*{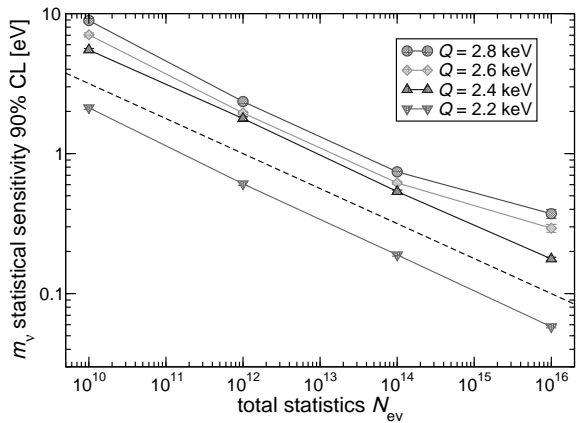}}
\end{center}
\caption{\label{fig:statistics}\Ho\ decay experiments statistical sensitivity dependence on the total statistics $N_{ev}$ for \de\fwhm$=1$\,eV, $f_{pp}=10^{-5}$, and $b=0$\,count/eV/s/detector.}
\end{figure}

Monte Carlo simulations confirm that the total statistics $N_{ev}$ is crucial to reach a sub-eV neutrino mass statistical sensitivity as for beta decay experiments (Figure\,\ref{fig:statistics}, see caption for the simulation details.)\cite{sensbeta}. In particular the sensitivity shows the same scaling as $N_{ev}^{-1/4}$ (dashed line in Figure\,\ref{fig:statistics}), as it would be naively expected for a $m_\nu^2$ sensitivity purely determined by statistical fluctuations. The uncertainty affecting the $Q$-value translates into about a factor 3 to 4 on the achievable neutrino mass sensitivity.

\subsubsection{Effect of experimental parameters}
\begin{figure}[!tb]
\begin{center}
\resizebox{0.42\textwidth}{!}{\includegraphics*[clip=true,trim=0pt 5pt 0pt 0pt]{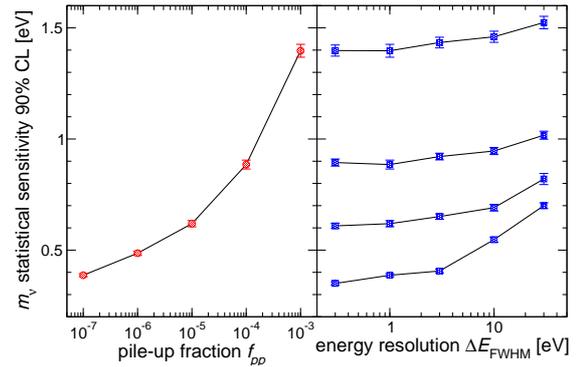}}
\end{center}
\caption{\label{fig:defpp}\Ho\ decay experiments statistical sensitivity dependence on pile-up fraction \fpp\ and energy resolution \de\fwhm\ for $Q=2600$\,eV, $N_{ev}=10^{14}$, and $b=0$\,count/eV/s/detector. Left: Energy resolution is fixed to \de\fwhm$=1$\,eV. Right: Pile-up fraction taken as (from bottom to top) $f_{pp}=10^{-7}$, $10^{-5}$, $10^{-4}$, and $10^{-3}$.}
\end{figure}

Figure\,\ref{fig:defpp} helps understanding the role of the detector performance in terms of time end energy resolutions.
Indeed the experimental sensitivity is not directly related to the time resolution, but only to the combination $f_{pp}=\tau_R\times A_\mathrm{EC}$, that is the amount of pile-up. In Figure\,\ref{fig:defpp} the sensitivity is plotted for constant energy resolution \de\fwhm\ (left) and for constant
\fpp\ (right), with the other experimental parameter varying in an interval of interest for typical detector configurations (see caption for more details). The plots suggest that the impact of the energy resolution on the sensitivity is relatively smaller than that of pile-up. Moreover, in presence of a high level of pile-up, the experiment is relatively less sensitive to the energy resolution. Qualitatively this latter effect can be understood in the following way: the more the pile-up hinders the signal at $Q$, the larger is the energy interval below $Q$ which weighs in the fit, and the less the energy resolution counts. 
However, it is worth noting that the time resolution depends on the detector signal-to-noise ratio at high frequency and therefore at constant bandwidth -- that is at constant detector rise time -- an energy resolution deterioration unavoidably turns in a worse time resolution. This effect it is not considered in the simulation.

\subsubsection{Trade-off between activity and pile-up}
\begin{figure}[!tb]
\begin{center}
\resizebox{0.45\textwidth}{!}{\includegraphics*[clip=true,trim=0pt 20pt 0pt 50pt]{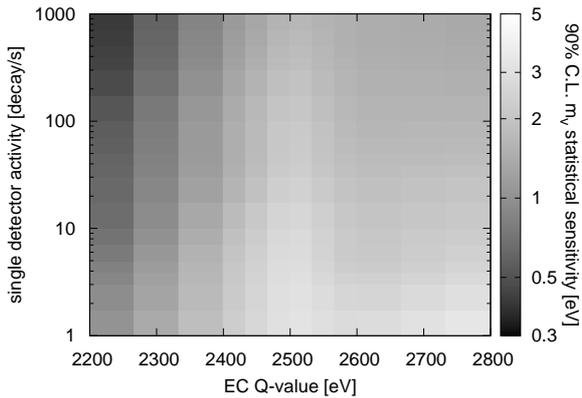}}
\end{center}
\caption{\label{fig:exposure}\Ho\ decay experiments statistical sensitivity dependence on $Q$-value and \Aec\ for \de\fwhm$=1$\,eV, $\tau_R=1\mu$s, $T=t_M\times N_{det}=3000$\,det$\times$year, and $b=0$\,count/eV/s/detector.}
\end{figure}

\begin{figure}[!hb]
\begin{center}
\resizebox{0.42\textwidth}{!}{\includegraphics*{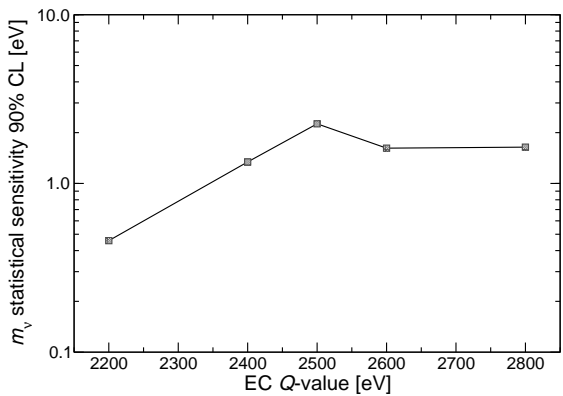}}
\end{center}
\caption{\label{fig:sensq}A slice of the data plotted in Figure\,\ref{fig:exposure} taken for \Aec\,$=100$\,decay/s.}
\end{figure}

Given the strong dependence of the sensitivity on the total statistics, for a fixed experimental exposure $T$ -- that is for a fixed measuring time and a fixed experiment size -- and for fixed detector performance, \de\fwhm\ and $\tau_R$, it always pays out to increase the single detector activity \Aec\ as high as technically possible, even at the expenses of an increasing pile-up level. This is exemplified in Figure\,\ref{fig:exposure} (see caption for the simulation details). Of course there maybe several limitations to the possible activity \Aec, such as, for example, the effect of the \Ho\ nuclei on the detector performance or  detector cross-talk and dead time considerations.
It is worth noting that in calorimetric measurements of the type considered here, in first approximation the   increase of single detector activity \Aec\ does not go along with an increase of the detector size (see \S\ref{calo}), which would translate in a performance degradation.
Figure\,\ref{fig:sensq} displays the statistical sensitivity achievable with a single detector activity of about 100\,decays/s under the same hypothesis for detector performance and exposure as in Figure\,\ref{fig:exposure}.

\begin{figure}[!htb]
\begin{center}
\resizebox{0.45\textwidth}{!}{\includegraphics*{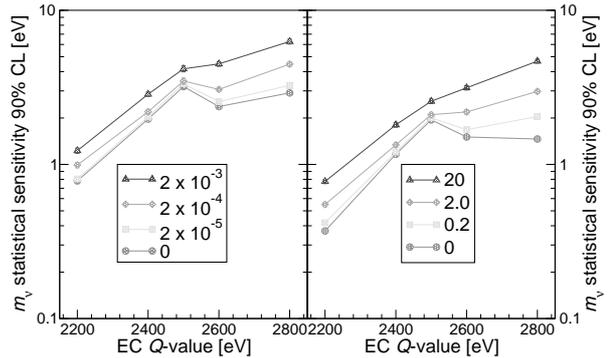}}
\end{center}
\caption{\label{fig:background}Effect of background on statistical sensitivity for $N_{ev}=10^{14}$ and \de\fwhm$=1$\,eV. Left: \Aec$=3$\,Hz/det and \fpp$= 3\times10^{-6}$, Right: \Aec$=300$\,Hz/det and \fpp$= 3\times10^{-3}$. The background levels in the boxes are in count/eV/day/detector units.}
\end{figure}
\subsubsection{Effect of background}
Because of the very low fraction of decays in the region of interest close to $Q$, the background  may be a critical issue in end-point neutrino
mass measurements.
The Monte Carlo simulations here are done with the hypothesis of a constant background $b$. A constant background is negligible as long as it is much smaller
than the pile-up spectrum, that is when $b \ll A_\mathrm{EC}f_{pp}/2Q$. Figure\,\ref{fig:background} confirms this simple considerations and shows that this is another good reason to have detectors with the highest possible activity. For large activities and correspondingly large pile-up rate, experiments should be relatively insensitive to cosmic rays and to environmental radioactivity. In a typical experiment with low temperature microcalorimenters, detectors have a sensitive area exposed to cosmic rays of the order of $10^{-8}$\,m$^2$ and a thickness of few micrometers: at sea level this translates in a cosmic ray interaction rate of about one per day with an average energy deposition of 10\,keV, which, in turns, means $b\lesssim10^{-4}$\,count/eV/day/detector. The flat background observed in the AgReO$_4$ microcalorimenters of the MIBETA experiment \cite{mibeta} 
was indeed measured to be about $1.5\times 10^{-4}$\,count/eV/day/detector, though comparison with \Ho\ decay experiments is difficult because of the different detector geometry and composition. All the above considerations should be complemented with an analysis of the effect of contaminations in the bulk of the detector -- especially $\beta$ and EC decaying isotopes -- and of the fluorescent X-ray and Auger emission from the material closely surrounding the detectors. The \Ho\ isotope production and its detector embedding are also likely to add radiactive contaminants internal to the detector: one example of dangerous isotope is $^{166m}$Ho ($\beta$ decay, $Q_\beta=1854$\,keV, $\tau_{1/2}=1200$\,year) which is produced together with \Ho\ in many of the production routes which have been proposed \cite{Enge12}.
A detailed analysis of the possible contaminations and their effects on the sensitivity is out the scope of the present work.
\begin{table*}[tpb]
\caption{Experimental exposures (in detector$\times$year) required for a neutrino mass statistical sensitivity of 0.1\,eV and with a single detector activity \Aec\ of 1\,decay/s. Different rows and columns are for different \de\fwhm\ (in eV) and \fpp\ values, respectively. The accuracy of the scaled values is at least 10\%.}
\label{tab:exposure}       
\begin{center}
\begin{tabular}{c|ccccc}
\hline\noalign{\smallskip}
$Q=2400$\,eV &	$10^{-7}$ &	$10^{-6}$ &	$10^{-5}$ &	$10^{-4}$ &	$10^{-3}$\\
\noalign{\smallskip}\hline\noalign{\smallskip}
0.3 	&$2.07\times 10^{8}$	&$7.37\times 10^{8}$	&$2.57\times 10^{9}$	&$9.50\times 10^{9}$	&$2.75\times 10^{10}$ \\
1   	&$2.41\times 10^{8}$	&$7.52\times 10^{8}$	&$2.57\times 10^{9}$	&$7.60\times 10^{9}$	&$2.39\times 10^{10}$ \\
3   	&$3.82\times 10^{8}$	&$7.07\times 10^{8}$	&$2.26\times 10^{9}$	&$7.91\times 10^{9}$	&$2.40\times 10^{10}$ \\
5   	&$4.36\times 10^{8}$	&$9.36\times 10^{8}$	&$2.44\times 10^{9}$	&$8.14\times 10^{9}$	&$2.48\times 10^{10}$ \\
\noalign{\smallskip}\hline\noalign{\smallskip}
$Q=2600$\,eV &	$10^{-7}$ &	$10^{-6}$ &	$10^{-5}$ &	$10^{-4}$ &	$10^{-3}$\\
\noalign{\smallskip}\hline\noalign{\smallskip}
0.3 	&$4.78\times 10^{8}$	&$1.57\times 10^{9}$	&$4.38\times 10^{9}$	&$2.03\times 10^{10}$	&$1.22\times 10^{11}$ \\
1   	&$7.14\times 10^{8}$	&$1.78\times 10^{9}$	&$4.68\times 10^{9}$	&$1.95\times 10^{10}$	&$1.21\times 10^{11}$ \\
3   	&$8.65\times 10^{8}$	&$1.75\times 10^{9}$	&$5.72\times 10^{9}$	&$2.29\times 10^{10}$	&$1.35\times 10^{11}$ \\
5   	&$1.19\times 10^{9}$	&$2.65\times 10^{9}$	&$6.27\times 10^{9}$	&$2.38\times 10^{10}$	&$1.39\times 10^{11}$ \\
\noalign{\smallskip}\hline\noalign{\smallskip}
$Q=2800$\,eV &	$10^{-7}$ &	$10^{-6}$ &	$10^{-5}$ &	$10^{-4}$ &	$10^{-3}$\\
\noalign{\smallskip}\hline\noalign{\smallskip}
 0.3 	&$  9.82\times10^{8}$	&$2.15\times10^{9}$	&$6.77\times10^{9}$	&$2.39\times10^{10}$	&$1.01\times10^{11}$ \\
1 	&$  1.43\times10^{9}$	&$2.73\times10^{9}$	&$7.60\times10^{9}$	&$2.59\times10^{10}$	&$1.04\times10^{11}$ \\
3 	&$  1.66\times10^{9}$	&$3.83\times10^{9}$	&$9.40\times10^{9}$	&$2.89\times10^{10}$	&$1.08\times10^{11}$ \\
5 	&$  2.83\times10^{9}$	&$4.47\times10^{9}$	&$1.06\times10^{10}$	&$3.06\times10^{10}$	&$1.11\times10^{11}$ \\
\noalign{\smallskip}\hline
\end{tabular}
\end{center}
\end{table*}

\subsubsection{Required experimental exposure}
Table\,\ref{tab:exposure} gives the exposure $T$ required for a target
neutrino mass statistical sensitivity of 0.1\,eV, for three $Q$-values, and for a range of meaningful experimental
parameters $\Delta E$ and $f_{pp}$. Exposures in the table are obtained by scaling the results of Monte Carlo simulations run for 
these parameter pairs and for a statistics of $10^{14}$\,decays.
Exposures are given for a single detector activity of 1\,decay/s and exposures for a different activity \Aec\  
can be obtained by simply dividing the given ones by the new $A^\prime_\mathrm{EC}$.
For a different target sensitivity $\Sigma^\prime_{90}(m_\nu)$, $T^\prime$ exposures can be obtained again by scaling $T$ in Table\,\ref{tab:exposure} as follows
\begin{equation}
\label{eq:expscal}
T^{\prime} = T \left[\frac{\Sigma_{90}(m_\nu)}{{\Sigma^\prime_{90}(m_\nu)}}\right]^4
\end{equation}
where $\Sigma_{90}(m_\nu)$ is the table target mass sensitivity.
For given pile-up fraction $f_{pp}$ and single detector activity  \Aec\ the corresponding detector time resolution
is obtained as $\tau_R=f_{pp}/A_\mathrm{EC}$.

\subsection{Conclusions}
In this paper I have discussed the statistical sensitivity of calorimetric \Ho\ electron capture neutrino mass experiments
using Monte Carlo simulations. Although assessing the real reach of this type of measurements requires also an extensive analysis of the systematic effects,
the results reported in this paper may be useful for designing the first generation of high statistics experiments aiming at sub-eV sensitivities.

\subsection{Aknowledgments}
I would like to thank for the useful discussions Joe Formaggio, Maurizio Lusignoli, Alvaro de Rujula, and the HOLMES collaboration members.%

\end{document}